\documentclass[11pt,letterpaper,english,epsf]{article}
\usepackage[T1]{fontenc}
\usepackage[utf8x]{inputenc}
\usepackage{verbatim}
\usepackage{amsmath}
\usepackage{amssymb}
\usepackage{setspace}
\usepackage{esint}
\makeatletter

\pdfpageheight\paperheight
\pdfpagewidth\paperwidth

\@ifundefined{definecolor}
 {\usepackage{color}}{}
\usepackage{amsfonts}\usepackage{amsfonts}
\usepackage{amsfonts}

\providecommand{\U}[1]{\protect\rule{.1in}{.1in}}

\def\b2hat{ {\hat b}_2 }

\def\D{\tilde{\nabla}}

\def\Cn{N^{(k)}_D}

\makeatother

\usepackage{babel}
\begin{document}

\title{Birkhoff's Theorem in Higher Derivative Theories of Gravity II. Asymptotically Lifshitz Black Holes}

\author{Julio Oliva, Sourya Ray\\
 \textit{Instituto de F\'{\i}sica, Facultad de Ciencias, Universidad Austral
de Chile, Valdivia, Chile.}\\
{\small julio.oliva@docentes.uach.cl, ray@uach.cl}}
\maketitle
\begin{abstract}
As a continuation of a previous work, here we examine the admittance
of Birkhoff's theorem in a class of higher derivative theories of
gravity. This class is contained in a larger class of theories which
are characterized by the property that the trace of the field equations
are of second order in the metric. The action representing these theories
are given by a sum of higher curvature terms. Moreover the terms of
a fixed order $k$ in the curvature are constructed by taking a complete
contraction of $k$ conformal tensors. The general spherically (hyperbolic
or plane) symmetric solution is then given by a static asymptotically
Lifshitz black hole with the dynamical exponent equal to the spacetime
dimensions. However, theories which are homogeneous in the curvature
(i.e., of fixed order $k$) possess additional symmetry which manifests
as an arbitrary conformal factor in the general solution. So, these
theories are analyzed separately and have been further divided into
two classes depending on the order and the spacetime dimensions.
\end{abstract}
\newpage{}

\section{Introduction}

In \cite{OR3}, the authors had presented a class of higher derivative
theories of gravity which admits Birkhoff's theorem in vaccum%
\footnote{This is a generalization of a theory cubic in curvature presented in \cite{OR1} to arbitrary
order.}. These theories belong to a bigger class of theories which are characterized
by second order traced field equations \cite{OR2}. We had shown that
there is a subclass of these theories whose field equations are generically
of fourth order but when evaluated on a spherically (hyperbolic or
plane) symmetric spacetimes reduce to second order thereby rendering
the admittance of Birkhoff's theorem i.e., the corresponding solution
is isometric to the static solution. Moreover the field equations
and the solutions have a similar structure to those of Lovelock theories
which are natural generalizations of Einstein's theory in higher dimensions
\cite{Lovelock}\footnote{Birkhoff's theorem in Lovelock gravity was proved in \cite{Zegers:2005vx}}. In the present work, as a continuation of the previous
one, we show that there is another subclass of this bigger class of
theories which admit Birkhoff's theorem. In this subclass, the scalar
invariants of a fixed order $k$ constituting the Lagrangian transform
covariantly under conformal resclalings of the metric. Such Lagrangians
are constructed by taking linear combination of complete contractions
of the conformal tensor
\begin{align}
C_{ab}^{\ \ cd}=R_{ab}^{\ \ cd}-\dfrac{4}{D-2}\delta_{[a}^{[c}R_{b]}^{d]}+\dfrac{2}{(D-2)(D-1)}\delta_{[a}^{[c}\delta_{b]}^{d]}R\ .
\end{align}
 We will call such scalars \textit{Weyl invariants} $W^{(k)}$, where
the \textit{order} $k$ is the number of conformal tensors constituting
the scalar. Then the action is expressed as a sum of terms, each of
a fixed order $k$ and is given by
\begin{align}
I^{(k)}=\int d^{D}x\sqrt{-g}\sum_{i}^{N_{D}^{(k)}}\alpha_{i}^{(k)}W_{i}^{(k)}\ ,\label{actionW}
\end{align}
 where $N_{D}^{(k)}$ is the number of independent Weyl invariants
$W_{i}^{(k)}$ of order $k$ in $D$ dimensions. Before we prove the
Birkhoff's theorem in the general class of theories given by an action
containing terms of different order $k$, we shall examine the field
equations and their solutions in theories with fixed order $k$. Such
an analysis requires to be classified into two separate cases depending
on the spacetime dimension $D$ and the order $k$. They are
\begin{itemize}
\item $D=2k$: In this case, the action is invariant under local conformal
transformations and hence the field equations and their solutions
have a conformal symmetry.
\item $D\neq2k$: In this case, the trace of the field equations is related
to the Lagrangian by
\begin{align}
\mathcal{G}_{a}^{(k)a}=\left(k-\dfrac{D}{2}\right)\mathcal{L}\ ,
\end{align}
 which implies that in vacuum the Lagrangian vanishes on its corresponding
solutions.
\end{itemize}
In the next section, we explicitly evaluate the field equations on
a spherically (hyperbolic or plane) symmetric spacetime ansatz. In
section 3, we show that the theories of fixed order $k$ admit a Birkhoff's
theorem in a slightly weaker sense. As stated before, the analysis
requires to be separated in two different cases depending on the spacetime
dimensions. In section 4, we take up the general non-homogeneous action
containing terms of different orders and prove the admittance of Birkhoff's
theorem where the corresponding solution is an asymptotically Lifshitz black hole. Finally, in section 5 we shall summarize our results and
their implications and mention some possible future directions of
study.

\section{Spherically (hyperbolic or plane) symmetric spacetimes}

Consider the general spherically (plane or hyperbolic) symmetric spacetimes
given by the following line element
\begin{align}
ds^{2}=\tilde{g}_{ij}(x)dx^{i}dx^{j}+e^{2\lambda(x)}d\Sigma_{\gamma}^{2},\label{ansatz}
\end{align}
 where $d\Sigma_{\gamma}^{2}=\hat{g}_{\alpha\beta}(y)dy^{\alpha}dy^{\beta}$
is the line element of a $(D-2)$-dimensional space of constant curvature
$\gamma$. Let $\D$ be the Levi-Civita connection on the two-dimensional
space orthogonal to the constant curvature space and $\tilde{R}$
be the corresponding scalar curvature. Then the nontrivial components
of the Riemann curvature tensor and the conformal tensor are given
by
\begin{align}
 & R_{jl}^{\ \ ik}=\frac{1}{2}\tilde{R}\delta_{jl}^{ik},\qquad\qquad\qquad\qquad R_{\nu\rho}^{\ \ \mu\lambda}=\tilde{\mathcal{B}}\delta_{\nu\rho}^{\mu\lambda},\qquad\qquad\qquad\qquad R_{j\nu}^{\ \ i\mu}=-\tilde{\mathcal{A}}_{j}^{i}\delta_{\nu}^{\mu},\\
 & C_{jl}^{\ \ ik}=\frac{(D-3)\tilde{S}}{2(D-1)}\delta_{jl}^{ik},\qquad C_{\nu\rho}^{\ \ \mu\lambda}=\frac{\tilde{S}}{(D-1)(D-2)}\delta_{\nu\rho}^{\mu\lambda},\qquad C_{j\nu}^{\ \ i\mu}=-\frac{(D-3)\tilde{S}}{2(D-1)(D-2)}\delta_{j}^{i}\delta_{\nu}^{\mu},
\end{align}
 where
\begin{align}
 & \tilde{\mathcal{B}}=\gamma e^{-2\lambda}-(\D_{m}\lambda)(\D^{m}\lambda),\\
 & \tilde{\mathcal{A}}_{j}^{i}=\D^{i}\D_{j}\lambda+(\D^{i}\lambda)(\D_{j}\lambda),\\
\text{and}\ \ \  & \tilde{S}=\tilde{R}+2\D_{k}\D^{k}\lambda+2\gamma e^{-2\lambda}.\label{s}
\end{align}
 %
Note that since all the components of the conformal tensor are a mere
multiple of the function $\tilde{S}$, each of the conformal densities
$W_{i}^{(k)}$'s evaluated on the metic (\ref{ansatz}) are proportional
to ${\tilde{S}}^{k}$. Let $W_{m}^{(k)}=\omega_{m}(D,k){\tilde{S}}^{k}$.
Then the field equations for the action (\ref{actionW}) evaluated
on the metric ansatz (\ref{ansatz}) are given by
\begin{align}
\mathcal{G}_{\ \ \ j}^{(k)i} & =k\left(\sum_{m=1}^{\Cn}\alpha_{m}^{(k)}\omega_{m}(D,k)\right)\tilde{\mathcal{P}}_{j}^{i}(\tilde{S}^{k-1})=k\alpha^{(k)}\tilde{\mathcal{P}}_{j}^{i}(\tilde{S}^{k-1}),\label{fe1}\\
\mathcal{G}_{\ \ \ \beta}^{(k)\alpha} & =k\left(\sum_{m=1}^{\Cn}\alpha_{m}^{(k)}\omega_{m}(D,k)\right){\delta}_{\beta}^{\alpha}\tilde{\mathcal{Q}}(\tilde{S}^{k-1})=k\alpha^{(k)}{\delta}_{\beta}^{\alpha}\tilde{\mathcal{Q}}(\tilde{S}^{k-1}),\label{fe2}\\
\mathcal{G}_{\ \ \ \alpha}^{(k)i} & =\mathcal{G}_{\ \ \ i}^{(k)\alpha}=0,\label{fe3}
\end{align}
 where $\tilde{\mathcal{P}}_{j}^{i}$ and $\tilde{\mathcal{Q}}$ are
two (related) second order linear differential operators defined on
the two-dimensional space orthogonal to the constant curvature base
manifold given by
\begin{align}
\tilde{\mathcal{P}}_{j}^{i} & =\Biggl[\delta_{j}^{i}\left(\frac{\tilde{R}}{2}+(D-1)\D_{k}\D^{k}\lambda+(D-2)(D-1)\D_{k}\lambda\D^{k}\lambda+\D_{k}\D^{k}+(2D-3)\D_{k}\lambda\D^{k}-\frac{\tilde{S}}{2k}\right)\nonumber \\
 & -(D-2)(\D^{i}\D_{j}\lambda+D\D^{i}\lambda\D_{j}\lambda)-\D^{i}\D_{j}-(D-1)(\D^{i}\lambda\D_{j}+\D_{j}\lambda\D^{i})\Biggr],\\
\tilde{\mathcal{Q}} & =-\frac{1}{D-2}\left[\tilde{\mathcal{P}}_{i}^{i}-\tilde{S}\left(1-\frac{D}{2k}\right)\right].\label{trrel}
\end{align}
 Note that (\ref{trrel}) implies that for $D=2k$, the trace of the
field equations vanish identically, as should be the case for any
conformally invariant theory.

\section{Birkhoff's theorem in theories of fixed order $k$}

We now prove a weaker version of Birkhoff's theorem for theories represented
by the action (\ref{actionW}). We present the proof for the cases
$D\neq2k$ and $D=2k$ separately. In the following, we will assume
\begin{equation}
 \alpha^{(k)}:=\sum_{m=1}^{\Cn}\alpha_{m}^{(k)}\omega_{m}(D,k) \neq 0.
\end{equation}
If $\alpha^{(k)}=0$, then any metric of the form (\ref{ansatz}) satisfies all the field equations.

\subsection{$D\neq2k$}

In this case, as explained earlier, the Lagrangian vanishes on its
corresponding solutions. When evaluated on the spherically symmetric
anstaz (\ref{ansatz}), this implies
\begin{equation}
\tilde{S}=\tilde{R}+2\D_{k}\D^{k}\lambda+2\gamma e^{-2\lambda}=0
\end{equation}
 Now, if we perform a conformal transformation in the $2$-dimensional
space as $\tilde{g}_{ij}\rightarrow e^{2\lambda(x)}\tilde{g}_{ij}$,
then the above equation takes the form $e^{-2\lambda}(\tilde{R}+2\gamma)=0$.
This in turn implies that the \textit{new} two dimensional metric
$\tilde{g}_{ij}$ is of constant curvature which always admits a non-null
Killing vector. Adapting to these coordinates the solution can be
written as
\begin{equation}
ds^{2}=e^{2\lambda\left(t,\rho\right)}\left[\Omega\left(\rho\right)\left(-dt^{2}+d\rho^{2}\right)+d\Sigma_{\gamma}^{2}\right]\ ,
\end{equation}
 where $\Omega(\rho)$ satisfies the equation
\begin{equation}
-\dfrac{1}{\Omega}\dfrac{d}{d\rho}\left(\dfrac{\Omega'}{\Omega}\right)+2\gamma=0\ .
\end{equation}
 which can be integrated to obtain
\begin{equation}
\Omega(\rho)=\begin{cases}C_{1}\cos^{-2}\left(\sqrt{C_{1}\gamma}(\rho+C_{2})\right) \qquad \text{when} \; \gamma\neq0\\
              C_2e^{C_1\rho} \qquad \qquad \qquad \qquad \qquad \; \text{when} \; \gamma=0
             \end{cases}
,
\end{equation}
 where $C_{1}$ and $C_{2}$ are integration costants. By a further
redefinition of the coordinates
\begin{equation}
\dfrac{1}{r}=\begin{cases}-\dfrac{b}{2\gamma}+\sqrt{\dfrac{C_{1}}{\gamma}}\tan\left(\sqrt{C_{1}\gamma}(\rho+C_{2})\right) \qquad \text{when} \; \gamma\neq0\\
              -b+\dfrac{C_2}{C_1}e^{C_1\rho} \qquad \qquad \qquad \qquad \qquad \quad\, \text{when} \; \gamma=0
             \end{cases}
\ ,
\end{equation}
 one can rewrite the metric in Schwarzschild-like coordinates as
\begin{equation}
ds^{2}=e^{2\tilde{\lambda}\left(t,r\right)}\left[-f\left(r\right)dt^{2}+\frac{dr^{2}}{f\left(r\right)}+r^{2}d\Sigma_{\gamma}^{2}\right]\ ,\label{dimext}
\end{equation}
 where $f(r)=ar^{2}+br+\gamma$ and the constants $a,b$ are related to $C_1$ and $C_2$..
This metric is conformally flat and may represent an asymptotically
locally flat or (A)dS black hole with a Cauchy horizon.

\subsection{$D=2k$}

In this case, as explained earlier, the action and the corresponding
field equations have a conformal symmetry. We exploit this symmetry
to set $\lambda=0$. Equation (\ref{fe1}) then gives
\begin{align}
\left[\delta_{j}^{i}\left(\frac{\tilde{R}}{2}+\Box-\frac{\tilde{S}}{2k}\right)-\tilde{\nabla}^{i}\tilde{\nabla}_{j}\right]\tilde{S}^{k-1}=0\ .\label{one}
\end{align}
 Note that equation (\ref{fe2}) is then manifestly satisfied. Taking
the trace of the above equation we obtain
\begin{align}
\left(\tilde{R}-\frac{\tilde{S}}{k}\right)\tilde{S}^{k-1}=-\tilde{\Box}\tilde{S}^{k-1}\;.\label{trace}
\end{align}
 Plugging this back into equation (\ref{one}) we get
\begin{equation}
\left(\delta_{j}^{i}\Box-2\tilde{\nabla}^{i}\tilde{\nabla}_{j}\right)\tilde{S}^{k-1}=0\;.\label{two}
\end{equation}
 Contracting the above equation by the two dimensional Levi-Civita
tensor $\epsilon_{ki}$ and then symmetrizing the indices $(j,k)$
we obtain
\begin{align}
\tilde{\nabla}_{(j}\epsilon_{k)i}\tilde{\nabla}^{i}\tilde{S}^{k-1}=0\ .
\end{align}
 This implies that the vector ${\tilde{\xi}}_{k}=\epsilon_{ki}\tilde{\nabla}^{i}\tilde{S}^{k-1}$
satisfies the Killing equation $\tilde{\nabla}_{(i}{\tilde{\xi}}_{k)}=0$.
We now show that if $\tilde{\xi}_{k}$ is a null Killing vector then
the two dimensional metic $\tilde{g}_{ij}$ is of constant curvature.
\begin{align}
 & {\tilde{\xi}}^{k}{\tilde{\xi}}_{k}=0\nonumber \\
\Rightarrow\quad & (\tilde{\nabla}^{i}\tilde{S})(\tilde{\nabla}_{i}\tilde{S})=0\nonumber \\
\Rightarrow\quad & (\tilde{\nabla}_{j}\tilde{\nabla}^{i}\tilde{S})(\tilde{\nabla}_{i}\tilde{S})=0\nonumber \\
\Rightarrow\quad & (\tilde{\Box}\tilde{S}^{k-1})(\tilde{\nabla}_{i}\tilde{S})=0\qquad\qquad\qquad\text{using (\ref{two})}\nonumber \\
\Rightarrow\quad & \tilde{\Box}\tilde{S}^{k-1}=0\nonumber \\
\Rightarrow\quad & \tilde{S}=0\quad\text{or}\quad\frac{2k\gamma}{k-1}\qquad\qquad\qquad\ \text{using (\ref{trace})}\ .
\end{align}
 Therefore, in case $\tilde{\xi}_{k}$ is a null Killing vector, then
the metric is of constant curvature which in turn implies that the
metric must admit at least one non-null Killing vector. So we could
again adapt to these coordinates and take the following metric ansatz
as previously
\begin{align}
ds^{2}=\Omega(\rho)\left[-dt^{2}+d\rho^{2}\right]+d\Sigma_{\gamma}^{2}\ .
\end{align}
 However, the field equations are not integrable in these coordinates.
So instead we choose the following metric ansatz in the Schwarzschild
coordinates
\begin{align}
ds^{2}=\frac{1}{r^{2}}\left[-f(r)dt^{2}+\frac{dr^{2}}{f(r)}\right]+d\Sigma_{\gamma}^{2}
\end{align}
 The Ricci scalar of the two-dimensional subspace is then given by
\begin{align}
\tilde{R}=-r^{3}\frac{d^{2}}{dr^{2}}\left(\frac{f(r)}{r}\right)\label{ricciscalar}
\end{align}
 Equation (\ref{two}) then gives
\begin{align}
\tilde{\nabla}^{t}\tilde{\nabla}_{t}\tilde{S}^{k-1}=\tilde{\nabla}^{\rho}\tilde{\nabla}_{\rho}\tilde{S}^{k-1}
\end{align}
 which can be integrated to obtain
\begin{align}
\tilde{S}=\tilde{R}+2\gamma=\left(\frac{c}{r}+d\right)^{\frac{1}{k-1}}\ .
\end{align}
 Assuming $c\neq0$, we can now use the expression (\ref{ricciscalar})
to further integrate the above equation and get
\begin{align}
f(r)=ar^{2}+br+\gamma-\frac{(k-1)^{2}}{c^{2}k(2k-1)}r^{2}\left(\frac{c}{r}+d\right)^{\frac{2k-1}{k-1}}\ .\label{f(r)}
\end{align}
 Substituting this in equation (\ref{trace}), we find a further constraint
among the integration constants given by
\begin{align}
bc=2\gamma d\ .
\end{align}
 Hence, redefining the constants $c$ and $b$, we can express the
metric function as
\begin{align}
f(r)=ar^{2}+2\gamma br+\gamma+cr^{2}\left(\frac{1}{r}+b\right)^{\frac{2k-1}{k-1}}\ .\label{fforcneq0}
\end{align}
The constant $c\neq0$ can be set to $1$ without any loss of generality. This can be seen by further redefining the constants as $c={c'}^{\frac{2k-1}{k-1}}$, $bc'=b'$ and $a=a'$ followed by the coordinate tranformations $r\rightarrow c'r$, $t\rightarrow t/c'$ and relabeling the curvature of the {\it new} $(D-2)$-dimensional space $\gamma'=\gamma/{c'}^2$ and finally removing all the primes. However, if $c=0$, then the general metric function satisfying the
field equations is given by
\begin{align}
f(r)=ar^{2}+br+d,\qquad\text{where}\qquad d=\gamma\ \ \text{or}\ \ -\frac{\gamma}{k-1}\ .\label{fforc=0}
\end{align}
 Therefore the general spherically symmetric solution of the theory
given by the action (\ref{actionW}) in dimensions $D=2k$ is given
by the metric (\ref{dimext}) with the function $f(r)$ given by (\ref{fforcneq0})
or (\ref{fforc=0}). The Birkhoff's theorem and the corresponding
solution for $k=2$ was obtained in \cite{Riegert} for $\gamma$=1
and in \cite{Klemm:1998kf} for arbitrary $\gamma$.

\section{Birkhoff's theorem for the general action}

Now we shall prove a Birkhoff's theorem when the action is an arbitrary
linear combination of terms of different orders. We also inlcude a
cosmological constant term. The action is then given by
\begin{align}
I=\int d^{D}x\sqrt{-g}\left(\alpha^{(0)}+\sum_{k}\sum_{i}^{N_{D}^{(k)}}\alpha_{i}^{(k)}W_{i}^{(k)}\right)\ .\label{actiongen}
\end{align}
 The field equations of the above action evaluated on the metric ansatz
(\ref{ansatz}) are then given by
\begin{align}
\mathcal{G}_{a}^{b}=-\frac{1}{2}\alpha^{(0)}\delta_{a}^{b}+\sum_{k}\mathcal{G}_{\ \ \ a}^{(k)b}\label{fe}
\end{align}
 Using the components of the field equations given by (\ref{fe1}),
(\ref{fe2}) and (\ref{trrel}) we then find the following polynomial
equation in $\tilde{S}$.
\begin{align}
\mathcal{G}_{a}^{a}=-\frac{D}{2}\alpha^{(0)}+\sum_{k}k\alpha^{(k)}\left(1-\frac{D}{2k}\right)\tilde{S}^{k}=0\ .\label{spoly}
\end{align}
 Let $\beta\neq0$ be a real root of the above polynomial. We next
choose the coordinates $(t,r)$ on the two-dimensional subspace such
that the coordinate $r=e^{\lambda}$ is spacelike and the metic (\ref{ansatz})
takes the form
\begin{align}
ds^{2}=-f(r,t)dt^{2}+\frac{dr^{2}}{g(r,t)}+r^{2}d\Sigma_{\gamma}^{2}\label{ansatz2}
\end{align}
 The $(t,r)$ component of the field equations then implies that $g=g(r)$.
Using this we solve the difference of $(t,t)$ and $(r,r)$ components
of the field equations. This implies
\begin{equation}
f(r,t)=\kappa(t)r^{2(D-1)}g(r)
\end{equation}
 where the arbitrary function $\kappa(t)$ can be absorbed by redefining
the coordinate $t$ and thus we can take $f=f(r)=r^{2(D-1)}g(r)$.
Next we use this and (\ref{s}) to integrate the equation $\tilde{S}=\beta$
and obtain the following form for the function $g(r)$.
\begin{align}
g(r)=\frac{a}{r^{2(D-2)}}+\frac{b}{r^{D-2}}-\frac{\beta}{2D(D-1)}r^{2}+\frac{\gamma}{(D-2)^{2}}
\end{align}
 where $a$ and $b$ are integration constants. Finally, we obtain
the sum of the $(t,t)$ and $(r,r)$ component of the field equations
which gives
\begin{align}
-\alpha^{(0)}+\sum_{k}k\alpha^{(k)}\beta^{k-1}\left[\frac{2}{D}\beta\left(1-\frac{D}{2k}\right)+(D-2)^{2}\frac{b}{r^{D}}\right]=0
\end{align}
 However, since $\beta$ is a solution of the equation (\ref{spoly}),
the above equation implies $b=0$ unless $\Sigma_{k}k\alpha^{(k)}\beta^{k-1}=0$.
Thus the general spherically (hyperbolic or plane) symmetric solution
of the theory given by the action (\ref{actiongen}) is
\begin{align}
ds^{2}=-\frac{r^{2D}}{l^{2D}}h(r)dt^{2}+\frac{dr^{2}}{\frac{r^{2}}{l^{2}}h(r)}+r^{2}d\Sigma_{\gamma}^{2}\label{lifshitz}
\end{align}
 where the function $h(r)$ is given by
\begin{align}
h(r) & =1+\frac{a}{r^{2(D-1)}}+\frac{\gamma l^{2}}{(D-2)^{2}r^{2}},\qquad\text{when }\sum_{k}k\alpha^{(k)}\beta^{k-1}\neq0\\
 & =1+\frac{a}{r^{2(D-1)}}+\frac{b}{r^{D}}+\frac{\gamma l^{2}}{(D-2)^{2}r^{2}},\qquad\text{when}\sum_{k}k\alpha^{(k)}\beta^{k-1}=0
\end{align}
 where $a$ and $b$ are new integration constants, $t$ has been
rescaled and $\beta=-\frac{2D(D-1)}{l^{2}}$ is a non-zero real solution
of (\ref{spoly}). Note that when $\alpha^{(0)}=0$ and $\beta=0$,
the general spherically (hyperbolic or plane) symmetric solution is
given by (\ref{dimext}). The metric (\ref{lifshitz}) is a static
asymptotically Lifshitz spacetime with the dynamical exponent equal
to the spacetime dimensions and represents a black hole for negative
values of $a$.

\section{Conclusions}

In summary, we have proved the admittance of Birkhoff's theorem in
a particular class of higher derivative theories. The action representing
these theories consists of invariants which are constructed by taking
complete contractions of a number of conformal tensors. We have shown
that in general the spherically (hyperbolic or plane) symmetric solution
of such a theory is given by an asymptotically Lifshitz spacetime
whose dynamical exponent is equal to the spacetime dimensions. However,
in the particular cases of theories homogeneous in the curvature,
there is an additional symmetry that manifests as an arbitrary conformal
factor in the general solution. These have been further classified
into two different cases based on the spacetime dimension $D$ and
the order $k$.

Let us now point out some of the important differences between the
theories considered in \cite{OR3} and here. In \cite{OR3} we had
considered a class of higher derivative theories whose field equations
when evaluated on spherically (hyperbolic or plane) symmetric spacetimes
reduce to second order. Moreover, the structure of these second order
equations is the same as those of Lovelock theories. In contrast,
here the theories under study yield field equations which after evaluating
on the spherically (hyperbolic or plane) symmetric ansatz are still
of fourth order. Consequently the general solution has a different
form than that of a Lovelock theory. This answers some of the questions
raised in \cite{OR3}. The analysis here explicitly shows that for
Birkhoff's theorem to hold it is not necessary that the field equations
evaluated on spherically (hyperbolic or plane) symmetric spacetimes
reduce to second order. Moreover, it shows that the admittance of
Birkhoff's theorem does not imply that the corresponding solutions
have the same form as those of Lovelock theories.

Now a few comments are in order. The homogeneous theories of order
$k$ in $D=2k$ dimensions are conformally invariant. The simlplest
case of $k=2$ in four dimensions is known as conformal gravity, originally
introduced by Bach in \cite{bach:1921} and has been a subject of
interest for various reasons. Recently it has been shown in \cite{Maldacena}
that by imposing a simple Neumann boundary condition on the metric,
conformal gravity can be shown to be equivalent to Einstein gravity
with a cosmological constant. This is possible because the solutions
of Einstein's gravity are also solutions of conformal gravity. It
is interesting to note that setting $a=b=0$ in (\ref{fforcneq0}) one
obtains the spherically symmetric solution of pure Lovelock theory
of order $k-1$.

\textbf{Acknowledgements}

We thank Eloy Ayón-Beato and Gaston Giribet for useful comments. This
work is supported by FONDECYT grants 11090281, 11110176 and CONICYT
791100027.

\end{document}